\documentclass[dvips]{acta}
\usepackage{supertabular,lscape,epsfig}
\usepackage{amssymb}
\usepackage{amsmath}
\SetPages{1}{9}
\SetVol{64}{2014}

\begin{document}

\begin{Titlepage}

\Title { TT Ari and its Quasi-Periodic Oscillations }

\Author {J.~~S m a k}
{N. Copernicus Astronomical Center, Polish Academy of Sciences,\\
Bartycka 18, 00-716 Warsaw, Poland\\
e-mail: jis@camk.edu.pl }

\Received{  }

\end{Titlepage}

\Abstract { Quasi-periodic oscillation (QPO) of TT Ari are transient, 
short-living phenomena. They appear and disappear and their periods and 
amplitudes vary on a time scale as short as 1 hour. 
Consequently the periodograms covering longer intervals of time are generally 
meaningless. 
}
{binaries: cataclysmic variables, stars: individual: TT Ari }

%Sec.1
\section { Introduction } 

TT Ari is a nova-like cataclysmic variable showing several types of variability 
(cf. Smak 2013 and references therein). Among them are: 
(1) negative superhumps with $P\approx 0.1329$d, and full amplitude  
$2A\approx 0.2$mag., often referred to as "3-hour" variations, and 
(2) transient, quasi-periodic oscillations (QPO's) with periods 
between 10 and 40 minutes and full amplitudes up to $2A\approx 0.2$mag. 

In spite of numerous investigations the evidence concerning the nature 
and characteristics of those QPO's is confusing:  

Williams (1966) from the analysis of one particular night found three 
QPO's being present in the second part of this run, but absent in its 
first part. 

Semeniuk et al. (1987) calculated global periodograms for several seasons and 
found just one persistent QPO, its period decreasing from $P_{QPO}\approx 27$ min 
in 1961/62 to $P_{QPO}\approx 17$ min in 1985.

Many authors (e.g. Andronov 1999, Kim 2009, Kraicheva et al. 1997,1999, 
Tremko et al. 1996, Udalski 1987) found several QPO's, with periods in the range 
$P_{QPO}\sim 10-40$ minutes, being often simultaneously present in the periodograms 
based on a single night or in global periodograms based on a given season. 

Vogt et al. (2013) analyzed the continuous 10-day light curve obtained 
in 2007 with the MOST satellite and found {\it no} QPO in the global periodogram 
in the range from 10 to 30 minutes and amplitude exceeding 0.25 percent. 

The purpose of the present paper is to clarify this situation by presenting  
results of a more detailed analysis of the behavior of individual QPO's.

%Sec.2
\section { The Data } 

The data used in the present investigation consist of 15 light curves 
(4 in V and 11 in U) collected by the author in 1961/62 at the Lick 
Observatory and in 1966 at the Observatoire de Haute Provence (see Smak 2013);  
they are identified in the first two columns of Table 1 below. 
The duration of those runs was from 2 to 5 hours. 

Prior to further analysis variations related to the negative superhumps 
were removed, using periods $P_{nSH}$ applicable to a given season and 
amplitudes $A_{nSH}$ applicable to a given night. 
Examples of such "pre-whitened" light curves were shown in the previous 
paper (Smak 2013, Figs.1 and 3).

%Sec.3
\section { The QPO Periods and Amplitudes } 

We begin by calculating periodograms. Each run is divided into two parts 
and three periodograms, covering periods from 5 to 60 minutes, are calculated 
separately for those two parts and for the entire run. 
In what follows we shall refer to them as "part 1", "part 2", 
and "both parts", with corresponding QPO periods and amplitudes being designated 
as $P(1)$, $P(2)$, $P(1+2)$ and $A(1)$, $A(2)$, $A(1+2)$. 

%***Fig.1
\begin{figure}[htb]
\epsfysize=12.0cm 
\hspace{1.0cm}
\epsfbox{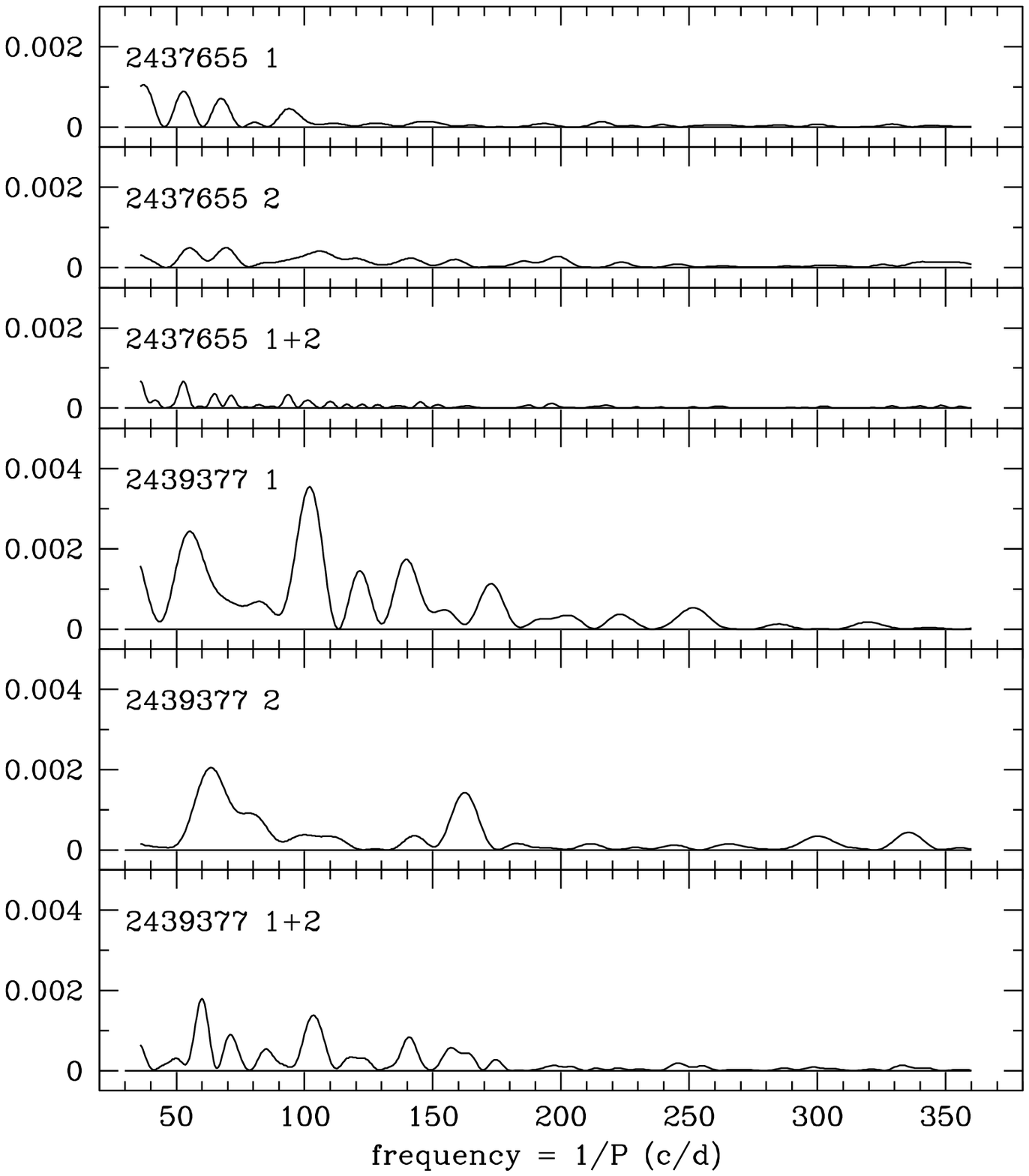} 
\vskip -0.5truecm
\FigCap { Examples of periodograms calculated separately for the two parts 
of a given run and for the entire run (1+2). 
}
\end{figure}

Examples of periodograms are shown in Fig.1 to illustrate their main  
characteristics: (1) periodograms for part 1, part 2, and both parts are 
generally different, and (2) different periodicities are present in 
periodograms obtained from different nights. 

To determine the QPO periods and amplitudes we proceed in the usual way: 
After finding the period and amplitude of the strongest QPO present 
in the periodogram we pre-whiten the light curve by subtracting this 
strongest signal and calculate the next periodogram. The procedure is 
repeated until no signal with amplitude exceeding of 0.03 mag. can be seen in the periodogram. 
This particular limit was set arbitrarily. Additional calculations showed, 
however, that making it lower would result in adding only few weaker 
periodicities not affecting our main conclusions. 

%*** Table 1
\begin{table}[h!]
{\parskip=0truept
\baselineskip=0pt {
\medskip
\centerline{Table 1}
\medskip
\centerline{ QPO Periods and Amplitudes }
\medskip
$$\offinterlineskip \tabskip=0pt
\vbox {\halign {\strut
\vrule width 0.5truemm #&	%1
\enskip\hfil#\hfil\enskip&      %2
\vrule#&			%3
\enskip#\enskip&                %4
\vrule#&			%5
\enskip\hfil#\hfil\enskip&      %6
\vrule#&			%7
\enskip\hfil#\hfil\enskip&      %8
\vrule#&			%9
\enskip\hfil#\hfil\enskip&      %10
\vrule#&			%11
\enskip\hfil#\hfil\enskip&      %12
\vrule#&			%13
\enskip\hfil#\hfil\enskip&      %14
\vrule#&			%15
\enskip\hfil#\hfil\enskip&      %16
\vrule width 0.5 truemm # \cr	%17
\noalign {\hrule height 0.5truemm}
& 2430000+ && C &&\hfil P(1)\hfil &&\hfil A(1)\hfil &&\hfil P(2)\hfil &&\hfil A(2)\hfil &&\hfil P(1+2)\hfil &&\hfil A(1+2)\hfil &\cr
\noalign {\hrule height 0.5truemm}
& 7646 && V &&{\bf 18.0} && 0.038 &&{\bf 19.6} && 0.032 && 19.1 && 0.033 &\cr
& 7655 && V && 38.8 && 0.032 && .... && ..... && .... && ..... &\cr
&      &&   && 21.1 && 0.030 && .... && ..... && .... && ..... &\cr
& 7656 && V && .... && ..... && 16.8 && 0.029 && .... && ..... &\cr
& 7660 && V && .... && ..... && 12.1 && 0.036 && .... && ..... &\cr
& 7660 && U && 16.2 && 0.048 && 13.4 && 0.040 && .... && ..... &\cr
&      &&   && .... && ..... && 41.1 && 0.040 && .... && ..... &\cr    %
& 7664 && U && 38.6 && 0.037 && 17.3 && 0.032 && .... && ..... &\cr
&      &&   && 20.4 && 0.031 && .... && ..... && .... && ..... &\cr
& 7672 && U && 50.0 && 0.054 && 16.1 && 0.083 && 15.4 && 0.062 &\cr    %
&      &&   && 29.0 && 0.050 && 37.2 && 0.070 && 24.4 && 0.059 &\cr
&      &&   && 20.7 && 0.037 &&  9.8 && 0.040 && .... && ..... &\cr
& 7675 && U &&{\bf 23.3}&& 0.070 && 57.3 && 0.051 && 25.7 && 0.040 &\cr   %
&      &&   && .... && ..... &&{\bf 21.9}&& 0.040 && 20.2 && 0.034 &\cr
& 7679 && U && .... && ..... && 17.5 && 0.041 && .... && ..... &\cr
&      &&   && .... && ..... && 14.1 && 0.037 && .... && ..... &\cr
&      &&   && .... && ..... && 37.9 && 0.034 && .... && ..... &\cr
& 7692 && U && 35.1 && 0.056 &&{\bf 13.9} && 0.043 && 19.4 && 0.035 &\cr
&      &&   &&{\bf 12.5} && 0.044 && 10.7 && 0.031 && 11.8 && 0.032 &\cr
&      &&   && .... && ..... && 20.1 && 0.030 && 29.7 && 0.031 &\cr
& 9360 && U && 26.8 && 0.051 && .... && ..... && 30.3 && 0.033 &\cr
&      &&   && 36.1 && 0.040 && .... && ..... && 25.7 && 0.030 &\cr
& 9375 && U && 38.0 && 0.054 && 18.5 && 0.054 && 18.5 && 0.040 &\cr
&      &&   && 26.6 && 0.038 && 22.5 && 0.044 && 24.3 && 0.037 &\cr
&      &&   && .... && ..... && 14.2 && 0.038 && 38.4 && 0.030 &\cr
&      &&   && .... && ..... && 11.7 && 0.030 && .... && ..... &\cr
& 9376 && U && 31.3 && 0.047 && 18.4 && 0.030 && 14.0 && 0.034 &\cr
&      &&   &&{\bf 13.6} && 0.043 &&{\bf 14.5} && 0.032 && .... && ..... &\cr
&      &&   && 43.5 && 0.032 && .... && ..... && .... && ..... &\cr    % 
& 9377 && U && 14.1 && 0.059 && 22.7 && 0.045 && 24.0 && 0.042 &\cr
&      &&   && 25.5 && 0.044 &&  8.9 && 0.037 && 13.9 && 0.037 &\cr
&      &&   && 53.1 && 0.039 && .... && ..... && .... && ..... &\cr    %
&      &&   && 10.4 && 0.034 && .... && ..... && .... && ..... &\cr
& 9378 && U &&{\bf 15.2} && 0.052 &&{\bf 15.4} && 0.046 && 19.9 && 0.044 &\cr
&      &&   &&{\bf 20.3} && 0.045 &&{\bf 19.1} && 0.046 && 15.8 && 0.043 &\cr
&      &&   && 28.5 && 0.036 && 48.8 && 0.032 && 14.6 && 0.032 &\cr    %
\noalign {\hrule height 0.5truemm}
}}$$
}}
\end{table}

%***Fig.2
\begin{figure}[htb]
\epsfysize=14.0cm 
\hspace{1.5cm}
\epsfbox{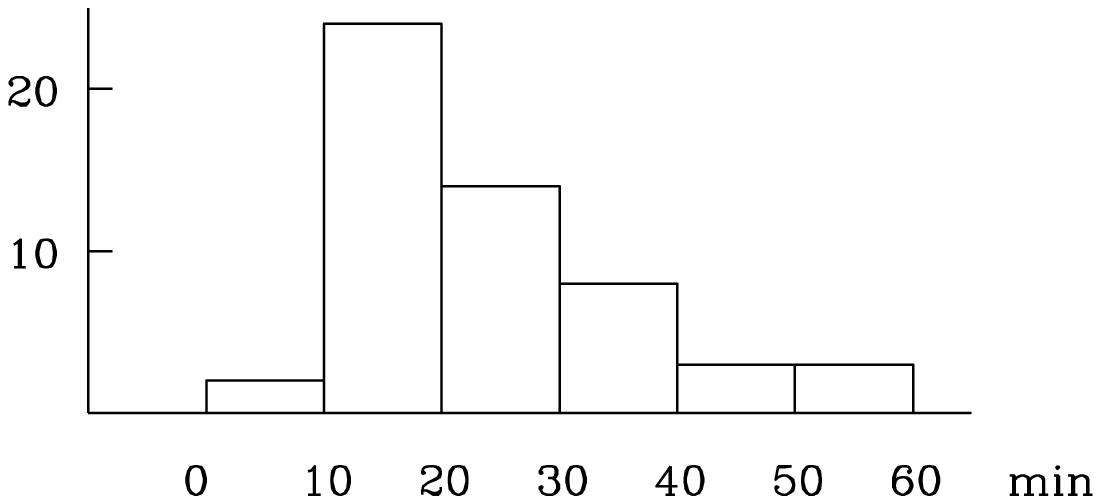} 
\vskip -9.5truecm
\FigCap { Histogram of QPO periods from part 1 and part 2 contained in Table 1. 
}
\end{figure}

Results are listed in Table 1, where periods are given in minutes and amplitudes 
-- in magnitudes, and the histogram of periods $P(1)$ and $P(2)$ 
contained in that table is shown in Fig.2. 
Those results can be summarized in the following points: 
(a) The QSO amplitudes in U are larger than in V. 
(b) The amplitudes $A(1+2)$ are lower than $A(1)$ and $A(2)$. 
(c) The QSO periods show concentration between 10 and 40 days. There are 
only two cases with periods shorter than 10 minutes (9.8 and 8.9 min.) and few 
periods longer than 40 minutes. 
(d) In about 50 percent of cases there are two or more QPO periods being 
simultaneously present in a given part. 
(e) The periods present in the two parts are generally different. 
There are only 6 cases with pairs of $P(1)$ and $P(2)$ (shown in Table 1 
in boldface) differing by less than 2 days. 
(f) There are only 11 cases with $P(1+2)$ being within 2 days of $P(1)$ or $P(2)$. 
(g) The 3 QPO periods obtained for the second part of JD 2437679 differ slightly 
from those obtained by Williams (1966), most likely due to the fact that 
he used incorrect value of $P_{nSH}$ ($0.1171$ instead of $0.1329$ d).
(h) No relation was found between the presence (or absence) of QPO's and 
the negative superhump phase, the orbital phase, or the beat phase. 

Using those results we can conclude that 
(1) the QPO's are short-living phenomena, 
(2) two or more QPO's can be simultaneously present, 
and (3) the periodicities present in periodograms obtained for "both parts" 
are either due to a strong QPO present in part 1 or part 2, or are artifacts 
unrelated to $P(1)$ or $P(2)$. 
We shall return to some of those points and strengthen those conclusions in the 
next Section.

%Sec.3
\section { The QPO Light Curves } 

We now turn to QPO light curves. For a specific QPO period $P_{QPO}$ 
the original light curve (pre-whitened with $P_{nSH}$; see Section 2) 
is pre-whitened with all other QPO periodicities. 
Then a series of {\it composite} light curves with $P_{QPO}$ are 
constructed, each of them including data points from 3 cycles, and each 
consecutive curve being shifted with respect to the previous one by one cycle. 
Two such series of composite light curves are shown, as examples, in Fig.3. 
The periods used in those two cases are the strongest periods detected 
in the periodograms: $P(2)=16.1$ min. in JD 2437672 and $P(1)=23.3$ min. 
in JD 2437675. 

As can be seen from Fig.3 the QPO amplitudes vary on a short time scale. 
In addition the curves are shifted in phase what is an obvious indication 
of period variations. To study those variations in more details 
we proceed as follows. For each light curve a cosine curve is fitted 
to the points giving the amplitude $A_{QSO}$ and the phase of maximum 
$\phi_{max}$. Those two parameters are then plotted as functions  
of time. Eight representative examples of the resulting 
$\phi_{max} vs.$ time and $A_{QSO}$ {\it vs.} time plots are shown 
in Figs.4-6 and discussed below. Note that the $\phi_{max} vs.$ time plots 
are equivalents of the $(O-C)$ diagrams. 

{\it JD 2437672} (Figs.3 and 4a). The QPO with $P(2)=16.1$ min. detected 
in part 2 was actually present also in part 1. Its amplitude was initially 
very low, increased on a time scale of $\sim 1$ hour, reaching maximum 
at JD 2437672.66 and begining to decrease afterwards. 
The period increased at a high rate: $dP/dt=+0.054\pm 0.018$. 

{\it JD 2437675} (Figs.3 and 4b). The QPO with $P(1)=23.3$ min. detected 
in part 1 had originally record high amplitude: $A\approx 0.11$ mag. 
It decrases rapidly, however, and after few cycles (or about 1 hour) 
it practically disappeared. 
The period decreased at a high rate: $dP/dt=-0.074\pm 0.014$. 

%*** Fig.3 i Fig.4 powinny byc na jednej stronie. 

%***Fig.3
\begin{figure}[htb]
\epsfysize=11.0cm 
\hspace{2.5cm}
\epsfbox{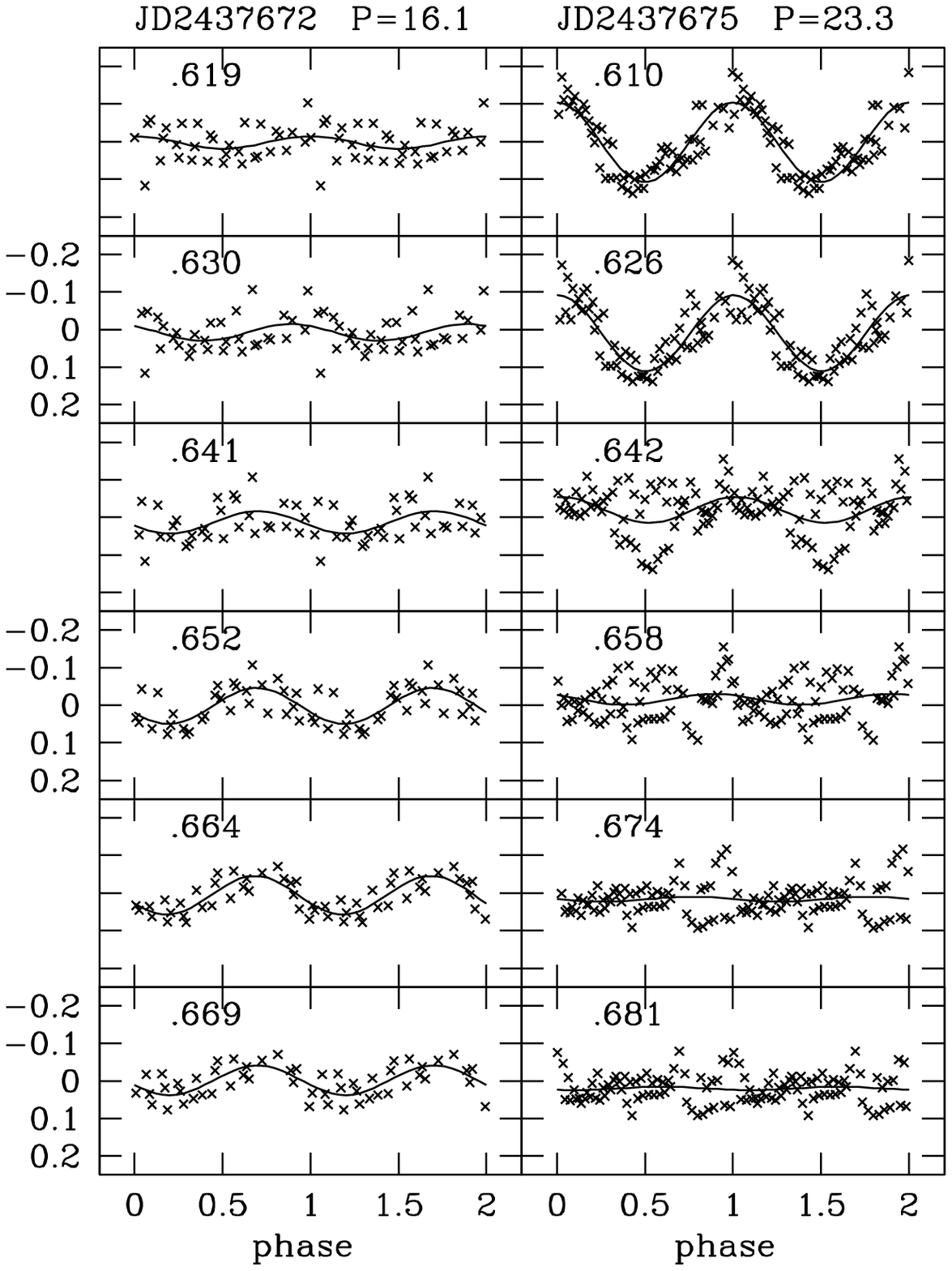} 
\vskip -1.5truecm
\FigCap { Composite light curves of QSO's with $P=16.1$ min on JD 2437672 
({\it left}) and $P=23.3$ min on JD 2437675 ({\it right}). 
Each curve includes 3 cycles and is shifted with respect to the previous one 
by one cycle. Solid lines represent the best fit cosine curves. 
See text for details. 
}
\end{figure}

%***Fig.4
\begin{figure}[htb]
\epsfysize=10.0cm 
\hspace{2.0cm}
\epsfbox{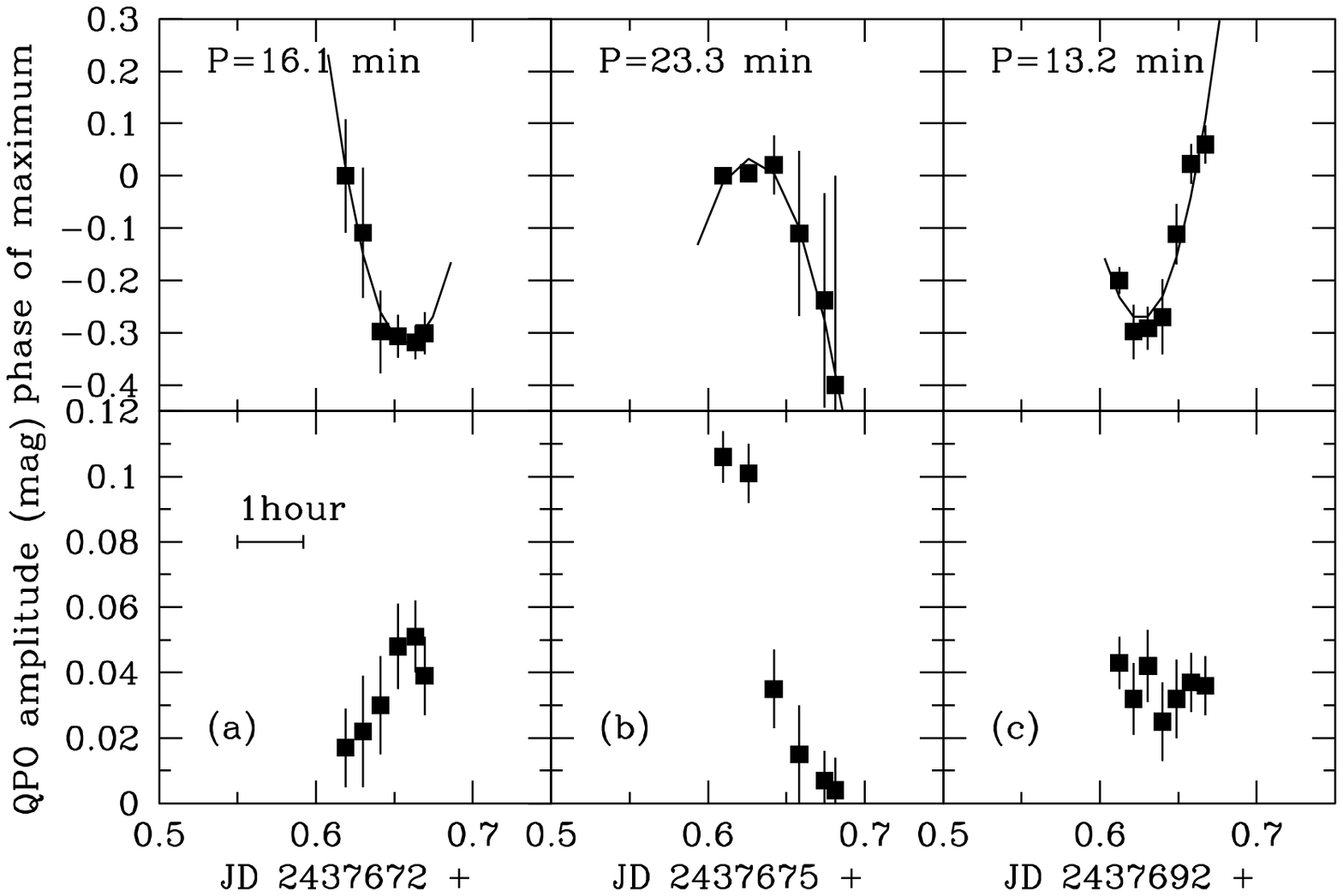} 
\vskip -3.0truecm
\FigCap { Variability of periods and amplitudes of three QPO's observed 
on JD 2437672, JD 2437675 and JD 2437692. Solid lines are the best fit 
parabolas describing period variations. See text for details. 
}
\end{figure}

%***Fig.5
\begin{figure}[htb]
\epsfysize=10.0cm 
\hspace{2.0cm}
\epsfbox{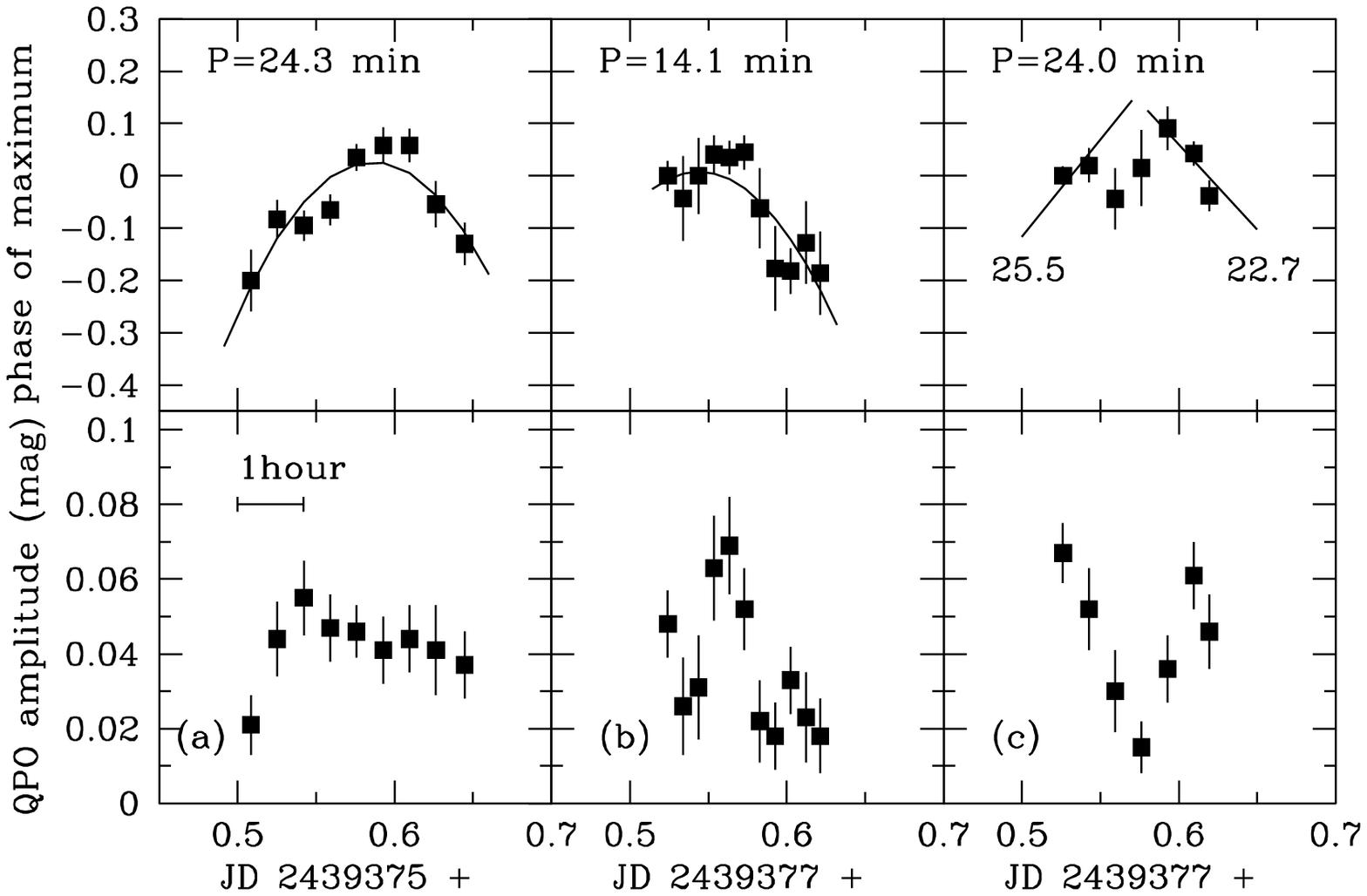} 
\vskip -3.0truecm
\FigCap { Variability of periods and amplitudes of three QPO's observed 
on JD 2439375 and JD 2439377. Solid lines represent the best fit cosine curves. 
Short lines in (c) represent periods identified in the periodograms. 
See text for details. 
}
\end{figure}

%***Fig.6
\begin{figure}[htb]
\epsfysize=10.0cm 
\hspace{3.5cm}
\epsfbox{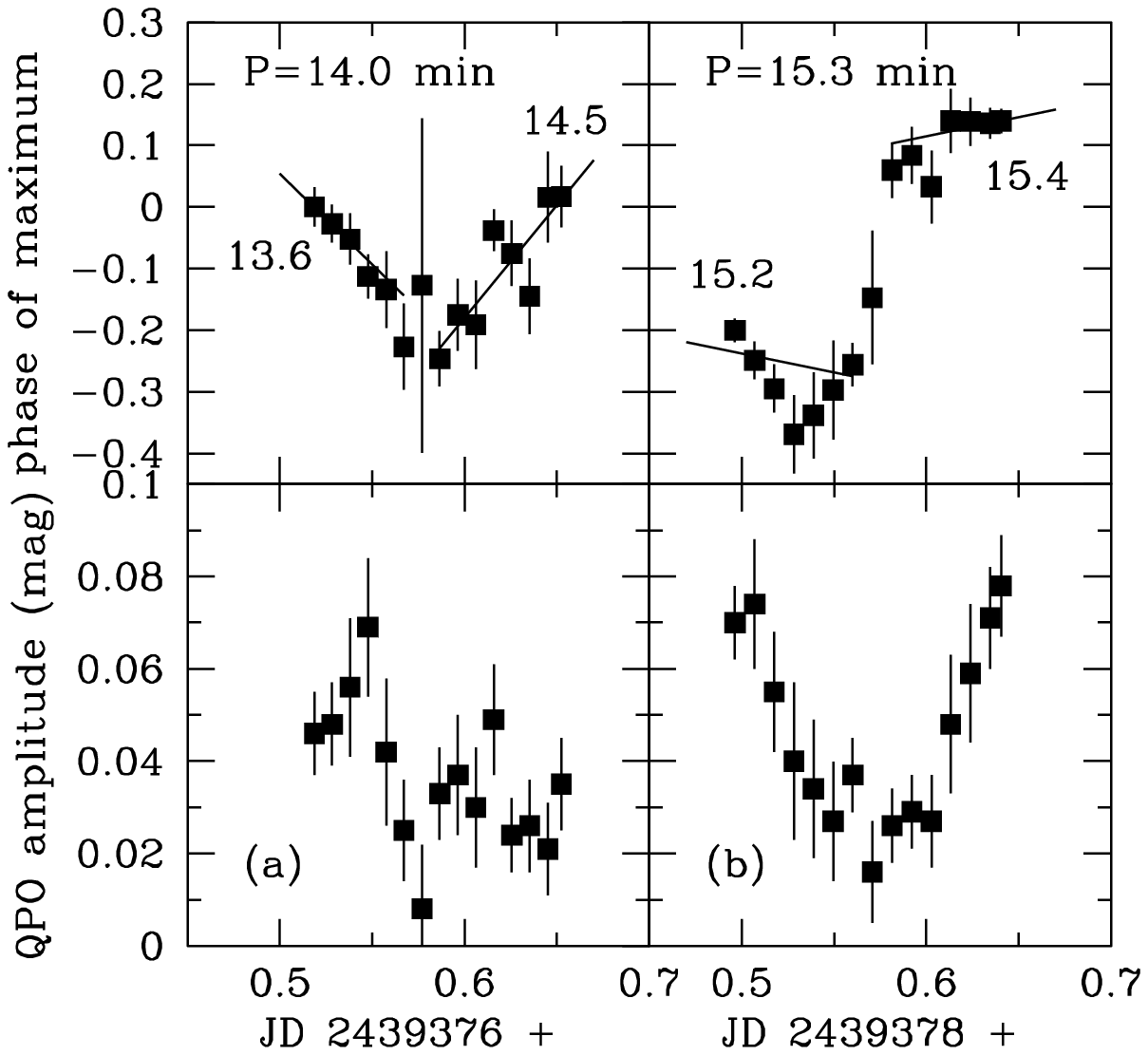} 
\vskip -3.0truecm
\FigCap { Variability of periods and amplitudes of QSO's observed on 
JD 2439376 and JD 2439378. Short lines represent periods identified 
in the periodograms. See text for details. 
}
\end{figure}

{\it JD 2437692} (Fig.4c). Two QPO's present in part 1 and part 2 had close 
periods: $P(1)=12.5$ and $P(2)=13.9$ min. Our analysis peformed 
with the mean value $P=13.2$ min shows that this was the same QPO 
with rapidly increasing period ($dP/dt=+0.038\pm 0.012$) and roughly constant 
amplitude. Regretably, this was the shortest run covering only 2 hours. 

{\it JD 2439375} (Fig.5a). Three periods detected in the periodogram: 
$P(1)=26.6$, $P(2)=22.5$, and $P(1+2)=24.3$ min. are close to $P=24$ min 
found by Semeniuk et al. (1987) in their global periodogram. 
Results of the analysis peformed with $P=24.3$ min show (Fig.5a) that 
this was one QPO with rapidly decreasing period ($dP/dt=-0.022\pm 0.005$). 

{\it JD 2439377} (Fig.5b). The QPO with $P(1)=14.1$ min. detected 
in part 1 was actually a short living feature with amplitude growing to 
maximum near JD 2439377.56, lasting for less than 1 hour, and then decreasing. 
The period decreased at a rate: $dP/dt=-0.007\pm 0.004$. 

{\it JD 2439377} (Fig.5c). Three periods detected in the periodogram: 
$P(1)=25.5$, $P(2)=22.7$, and $P(1+2)=24.0$ min. are close to $P=24$ min 
detected by Semeniuk et al. (1987) in their global periodogram. 
Results of the analysis peformed with $P=24.0$ min show (Fig.5c) that 
there were two different QPO's. The first, with $P=25.5$, rapidly declined 
in amplitude and near JD 2439377.57 was replaced by another one 
with $P(2)=22.7$ and rapidly growing amplitude. Therefore the period 
$P=24$ min was an artifact unrelated to the two real periodicities. 

{\it JD 2439376} (Fig.6a). Two QPO present in part 1 and part 2 had close 
periods: $P(1)=13.6$ and $P(2)=14.5$ min. Results of the analysis peformed 
with the mean value $P=14.0$ min (Fig.6a) could -- at first sight -- suggest 
that this was the same QPO with rapidly increasing period. One should note, 
however, that the amplitude, after reaching maximum near JD 2439376.55, 
declined rapidly and around JD 2439376.575 the QPO with $P=13.6$ disappeared,  
being replaced by another one with $P(2)=14.5$. The two solid lines represent 
periods $P(1)=13.6$ and $P(2)=14.5$ min. Note that they were nearly constant.  

{\it JD 2439378} (Fig.6b). This is another example of two strong QPO present 
in part 1 and part 2 having close periods: $P(1)=15.2$ and $P(2)=15.4$ min. 
Results of the analysis peformed with the mean value $P=15.3$ min (Fig.6b) 
are unambiguous: A large jump in $\phi_{max}$ clearly shows that there were 
two different QPO's. The first QPO with $P=15.2$ rapidly declined in amplitude 
and near JD 2439378.57 was replaced by another one with $P(2)=15.4$ and rapidly 
growing amplitude. The two solid lines represent periods $P(1)=15.2$ and 
$P(2)=15.4$ min. Note that they were variable: $P(1)$ -- increasing and 
$P(2)$ -- decreasing. 

%***Fig.7
\vskip -1.0truecm
\begin{figure}[htb]
\epsfysize=13.0cm 
\hspace{0.0cm}
\epsfbox{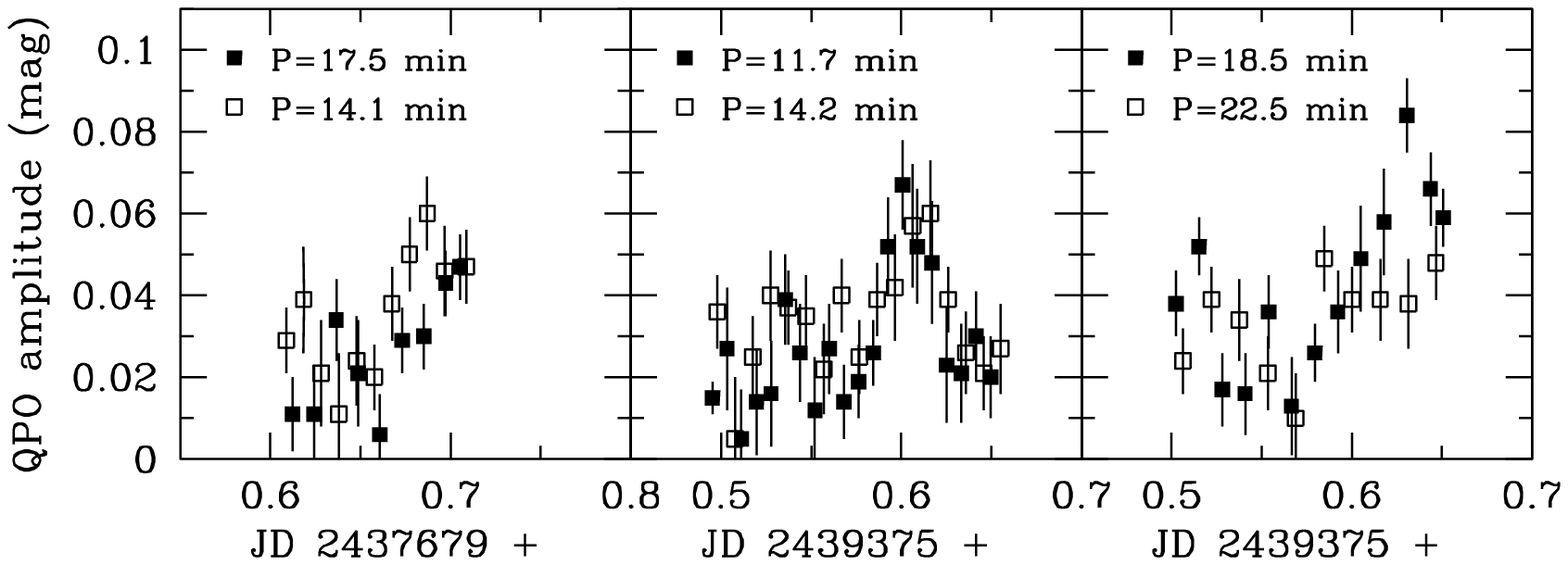} 
\vskip -7.5truecm
\FigCap { Variability of QSO amplitudes observed simultaneously on 
JD 2437679 and JD 2439375.  
}
\end{figure}

The $A_{QSO}$ {\it vs.} time plots permit also to check whether multiple 
QPO's identified in the periodograms were indeed present simultaneously. 
Fig.7 shows the amplitudes of two QPO's detected in part 2 
of JD 2437679 and of four QPO's in part 2 of JD 2439375. 
In both cases the QPO amplitudes varied in a similar way, on a time scale 
of $\sim$1 hour, reaching their maxima simultaneously or nearly 
simultaneously.

%Sec.4
\section { Discussion } 

The most important result of the present investigation is that the quasi 
periodic oscillations of TT Ari are transient, short-living phenomena which 
appear and disappear on a time scale as short as 1 hour. Their amplitudes 
and periods are strongly variable on a similar, very short time scale. 

The obvious consequence of this behavior is that periodograms calculated 
from data covering longer intervals of time, particularly global periodograms 
covering the entire season, only seldom can show real periodicities, but -- 
generally -- are meaningless. 
This is best illustrated by the global periodogram obtained by Vogt et al. 
(2013, Fig.3) from data covering 10 days: it did not show {\it any} QPO's with 
amplitude exceeding 0.0025 mag. On the other hand, however, their light curves 
(Vogt et al. 2013, Fig.1) showed clearly many transient QPO's with full 
amplitudes $2A$ up to 0.1 mag. (for example, there was a strong QPO 
with $P\approx 23$ min on JD 2454404).  

The nature and origin of QPO's in TT Arietis remain unclear...

\begin {references} 

\refitem {Andronov, I.L., Arai, K., Chinarova, L.L., Dorokhov, N.I., Dorokhova, T.N., 
      Dumitrescu, A., Nogami, D., Kolesnikov, S.V., Lepardo, A., Mason, P.A., 
      Matsumoto, K., Oprescu, G., Pajdosz, G., Passuelo, R., Patkos, L., Senio, D.S., 
      Sostero, G., Suleimanov, V.F., Tremko, J., Zhukov, G.V., Zo{\l}a, S.} 
      {1999} {\AJ} {117} {574} 

\refitem {Kim, Y., Andronov, I.L., Cha, S.L., Chinarova, L.L., Yoon, J.N.} 
     {2009} {\AA} {496} {765}

\refitem {Kraicheva, Z., Stanishev, V., Iliev, L., Antonov, A., Genkov, V.} 
      {1997} {\AA Suppl.Ser.} {122} {123} 

\refitem {Kraicheva, Z., Stanishev, V., Genkov, V., Iliev, L.} 
      {1999} {\AA} {351} {607}

\refitem {Semeniuk, I., Schwarzenberg-Czerny, A., Duerbeck, H., Hoffman, M., 
     Smak, J., St{\c e}pie{\' n}, K., Tremko, J.} {1987} {\Acta} {37} {197}

\refitem {Smak, J.} {2013} {\Acta} {63} {453} 

\refitem {Tremko, J., Andronov, I.L., Chinarova, L.L., Kumsiashvili, M.I., 
      Luthardt, R., Pajdosz, G., Patk{\'o}sz, L., R{\"o}{\ss}inger, S., Zo{\l}a, S.}
      {1996} {\AA} {312} {121}

\refitem {Udalski, A.} {1988} {\Acta} {38} {315}

\refitem {Vogt., N., Chen{\'e}, A.-N., Moffat, A.F.J., Matthews, J.M., Kuschnig, R., 
      Guenther, D.B., Rowe, J.F., Ruci{\'n}ski, S., Sasselov, D., Weiss, W.W.} 
      {2013} {\it Astronomische Nachrichten} {334} {1101}

\refitem {Williams, J.O.} {1966} {\PASP} {78} {279} 

\end {references}

\end{document}